# Unveiling Coverage-Dependent Interactions of N-Methylaniline with the Pt(111) Surface

Bushra Ashraf,[§] Nils Brinkmann,[§] Dave Austin, Duy Le, Katharina Al-Shamery,* and Talat S. Rahman*

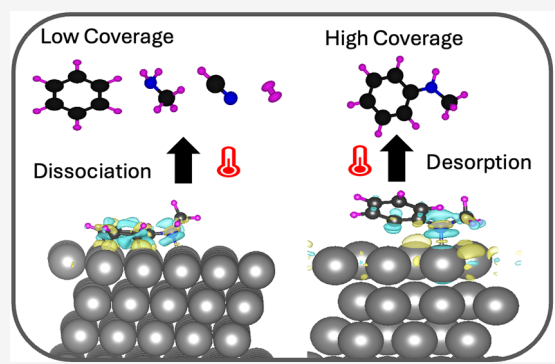

**ABSTRACT:** This study aims to elucidate the adsorption and surface chemistry of N-methylaniline (NMA) on Pt(111), using it as a model molecule to probe the activation mechanisms of aromatic amines on catalytic surfaces. Through a combination of density functional theory (DFT) calculations and experimental techniques such as temperature-programmed X-ray photoelectron spectroscopy (TP-XPS), temperature-programmed desorption (TPD), and Fourier transform infrared reflection absorption spectroscopy (FT-IRRAS), we explored the coverage-dependent behavior of NMA on Pt(111) to identify key steps in the activation process. The population of certain reaction paths is driven by a coverage-dependent balance between molecule surface charge transfer and intermolecular interactions, dictating the selective activation of specific bonds. Our findings reveal how coverage influences the orientation and bonding of NMA on the Pt(111) surface. At lower coverages, the molecule binds to the surface through the phenyl ring and activation, facilitating C–N bond cleavage to the ring under HCN formation. In comparison, at higher coverages, the molecule binds only through the nitrogen atom and desorbs intact. These insights into variable bond activation lay the groundwork for understanding the fundamental processes involved in potential heterogeneously catalyzed reactions of aromatic amines, contributing to the development of new catalytic strategies.

## 1. INTRODUCTION

The synthesis of amines represents a cornerstone of the industrial market, covering everyday applications such as pharmaceuticals, agricultural chemicals, and polymers, accounting for billions of dollars in revenue.[1] The industrial recipe for the synthesis of amines hinges on factors such as the composition of the desired amine, the generated byproducts, and the availability of raw materials. For instance, methylamines are primarily manufactured through the reaction of methanol and ammonia utilizing supported zeolite catalysts, while catalyzed reductive amination finds extensive application in producing various alkylamines.[1,2] The latter process involves the reaction between ketones and aldehydes with ammonia, amines, or nitro compounds in the presence of hydrogen, often facilitated by supported metal catalysts.[1,2]

In amine chemistry, homogeneous catalysts offer promising selectivity but introduce the challenge of catalyst separation, which is particularly critical for drug production and requires high levels of purity. On the other hand, heterogeneous catalysts face challenges of product selectivity, including overalkylation and surface deactivation because of coking. The formation of byproducts presents cost-intensive issues, necessitating additional process steps for byproduct separation or catalyst regeneration, resulting in increased equipment and energy requirements. This underscores the critical importance of developing new selective catalysts, synthesis routes, and methodologies[3,4] for the synthesis of amines, possibly through heterogeneous catalysis. A fundamental understanding of the interaction between amines and metal surfaces is the key to developing these new synthesis routes. As a first step, amine adsorption characteristics may provide insights as amine–surface interactions may give rise to new surface species besides the molecularly adsorbed amines.[4,5] A comprehensive understanding of the activation mechanisms of both aliphatic and aromatic amines is thus imperative for advancing this field.

Platinum (Pt), renowned for its high reactivity as a transition metal catalyst, provides an excellent substrate for modeling surface properties, particularly on its (111) surface. This surface stands out because of its symmetry and densely packed structure.[6,7] In amine chemistry, a pivotal step involves activation of the N–H bond. An ideal model system for examination of such activation could be the interaction of the C atoms of the phenyl ring in NMA with Pt(111), as it allows insight into interactions between the aromatic ring and the possible activation of the N−H or C−N bond with the electron lone pair at the nitrogen atom that is in contact with the Pt surface. In fact, earlier investigations on the Pt(111) surface have unveiled a significant occurrence of N−H cleavage activation.[8−10]

The adsorption of molecules on metal surfaces is a complex interplay of factors encompassing interactions between the adsorbate and the surface and intermolecular interactions among neighboring adsorbate molecules. These interactions can result in net attractive or repulsive forces. Comprehensive inquiries that delve into the surface chemistry of methylamines when exposed to the Pt(111) surfaces have enhanced our grasp of amine−surface interactions, revealing the formation of aminocarbyne species as intermediates of methylamine on Pt(111).[6,10,11] These studies have consistently indicated that amine adsorption occurs using the nitrogen lone pair. Subsequent decomposition occurs through the desorption of hydrogen cyanide and hydrogen at temperatures exceeding 400 and 500 K. Interestingly, extending the aliphatic chain, as seen with ethylamine, appears to yield analogous amino-vinylidene species, akin to the shorter aminocarbyne species formed by methylamine.[12] An intriguing question arises concerning how the amine−surface interaction differs for aromatic amines. It is known that aromatic compounds like benzene, toluene, or xylene tend to adsorb with the phenyl ring oriented parallel or nearly parallel to the surface, forming $\pi$-bonds with the metal surface.[13,14] At high coverages or under decomposition conditions, the ring can experience tilting.[15,16] Aromatic amines, such as aniline and its derivatives, such as aminophenol, hold particular interest because of their propensity to form linear polyaniline chains on Pt(111) within a specific temperature range from 475 to 495 K, and remain stable up to 560 to 700 K.[17−19] Because of their remarkable stability and electrical conductivity,[20] polyanilines have garnered substantial attention for polymer applications.

Here, we investigate the adsorption and decomposition behaviors of one such aniline, N-methylaniline (NMA) on Pt(111), through an integrated approach that involves ab initio density functional theory (DFT)-based calculations intertwined with experimental findings. Our primary focus is understanding the coverage-dependent interactions of the various functional components within NMA and with the Pt(111) surface. Since NMA features an aromatic ring and represents a secondary amine with distinct substituents linked to the nitrogen atom, it offers an excellent opportunity to probe the potential activation of C−H bonds in the methyl group, C−H bonds within the ring, the N−H bond in the amine group, or the C−N bonds connecting the ring and amine and the methyl group and amine. Additionally, this work addresses whether electron donation occurs through the lone pair of the amine or via $\pi$-interactions involving the aromatic ring with the metal surface. The insights from the DFT calculations on the electronic and geometric structure and the nature of the bonding and charge transfer, complemented by experimental findings garnered from techniques such as temperature-programmed desorption (TPD), X-ray photoelectron spectroscopy (XPS), and Fourier transform infrared reflection−absorption spectroscopy (FT-IRRAS), are poised to shed light on the intricacies of the surface chemistry of this particular amine on Pt(111). This paper is further organized as follows: Section 2 provides a detailed account of the computational and experimental methodologies employed. The results are presented in Section 3, which is further divided into theoretical (Section 3.1) and experimental (Section 3.2) subsections. Section 4 offers a comprehensive discussion of these findings, while Section 5 concludes the paper by summarizing the key insights drawn from the study.

## 2. COMPUTATIONAL AND EXPERIMENTAL DETAILS

The DFT calculations were performed using the Quantum Espresso package,[20] which performs an iterative solution of the Kohn−Sham equations in a plane-wave basis set. Plane waves with a kinetic energy cutoff of 500 Ry were used in the calculation. The exchange-correlation energy was calculated within the generalized gradient approximation (GGA) using the functional proposed by Perdew−Burke−Ernzerhof (PBE)[21] with Grimme's DFT-D3 van der Waals correction.[22] The electron−ion interactions for C, H, N, and Pt were described by using the projector-augmented wave (PAW) method to describe ionic cores. This is essentially an all-electron frozen-core method, combining the accuracy of all-electron methods and the computational simplicity of the pseudopotential approach. A Gaussian smearing function with a width of 0.1 eV was used to account for fractional occupancies. The optimized lattice parameter for bulk Pt was found to be 3.93 Å (the experimental value[23] is 3.92 Å), calculated for the face-centered cubic (fcc) crystal structure, and its reciprocal space was sampled with a 16 × 16 × 16 k-point grid generated automatically using the Monkhorst−Pack method.[24] The k-point mesh has been tested for each coverage such that the energy converged to 1 meV. The optimized k-mesh for various coverages are as follows: 5 × 7 × 1 for 1/6 coverage, 5 × 5 × 1 for 1/9 coverage, 4 × 4 × 1 for 1/16 coverage, 3 × 3 × 1 for 1/25 coverage, and 2 × 2 × 1 for 1/36 coverage. The calculated Pt band structure verifies the metallic nature of bulk Pt (Figure S1). To model the Pt(111) surface, a slab with a thickness between 4 and 7 layers was tested. Surface energy converged for a 5-layer model system with the bottom two layers fixed at bulk coordinates. A vacuum layer of 15 Å was used to minimize the interaction of adjacent unit cells along the z-axis. Geometry optimizations were stopped when all forces acting on atoms were less than 0.001 Ry/Bohr. For the characterization of the Projected Density of States (PDOS), a mesh of 11 × 11 × 1 was used in all cases. Vibrational frequencies were computed by using the harmonic approximation. The Hessian (second derivative) matrix was obtained numerically by independently displacing the nearest-neighbor atoms by 0.01 Å from their equilibrium positions. Zero-point energy correction, calculated using the Phonopy code,[25] has been added to the binding energy for each case studied.

Measurements for X-ray photoelectron spectroscopy (XPS) were carried out in a UHV chamber different from that used for temperature-programmed desorption (TPD) and Fourier transform infrared reflection absorption spectroscopy (FT-IRRAS) studies. The UHV chambers (base pressure below $10^{-10}$ mbar) were connected via a parking station to transfer the sample without breaking the vacuum. These chambers were equipped with a liquid nitrogen-cooled manipulator, a pinhole dosing system for organic compounds, and an ion source with a high-purity gas supply (argon, 99.999%, Air Liquide) connected through a leak valve for sputtering. A K-type thermocouple (CHAL-005, Omega Engineering) was spot-welded to the crystal to monitor its temperature. The crystal was mounted in a home-built sample holder, whose description can be found elsewhere.[26]

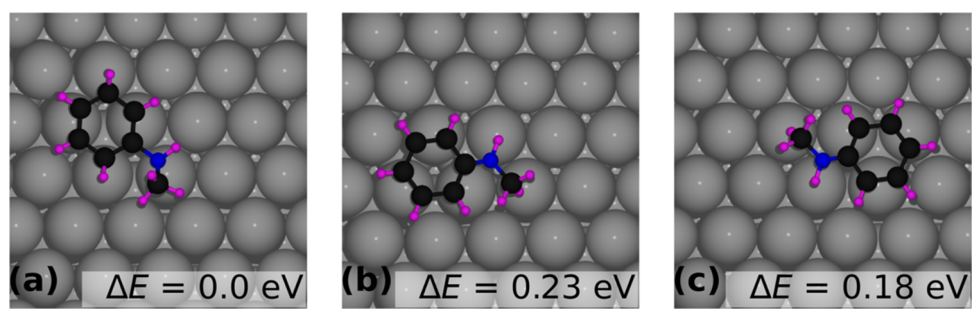

**Figure 1.** Adsorbed configuration of NMA on the Pt surface (a) bridge site, (b) fcc site, and (c) hcp site labeled with the relative energy difference of (b, c) configuration with (a).

A commercial Pt(111) single crystal (MaTeck, 10 mm diameter and 1 mm thickness) was used for these studies. The crystal was cleaned by sputtering with argon ions for 15 min at 298 K at $5 \times 10^{-5}$ mbar argon pressure and subsequent annealing for 10 min at 900 K. The cleanliness of the Pt(111) surface was checked with XPS and TPD for possibly remaining carbon residues. The Pt(111) long-range surface structure was confirmed by low-energy electron diffraction (LEED). For the adsorption of N-Methylaniline (NMA) (supplied by Thermofisher Scientific, 99%) on Pt(111), the sample was cooled to liquid nitrogen temperature and positioned a few millimeters in front of the pinhole doser. With a pressure of 1 mbar behind the pinhole, NMA was dosed for different dosing times after the butterfly valve was opened to the dosing compartment. TPD spectra were taken for different coverages of NMA on Pt(111). For dosing in the XPS chamber, a pressure of $1.5 \times 10^{-2}$ mbar was chosen in the pinhole doser. Because of the different sizes of the pinholes, the dosing times and pressures in the two chambers are not equivalent.

Temperature-programmed desorption spectra were measured with a quadrupole mass spectrometer (QMS Pfeiffer, Vacuum Prisma QMA 200) equipped with a Feulner cup (diameter of the nozzle around 5 mm) and channeltron detector. The sample was positioned a few mm in front of the Feulner cup and heated from 120 to 900 K with a heating ramp of 2 Ks$^{-1}$. The mass spectra were recorded with a 50 ms dwell time. The Redhead method[27] was used to determine the desorption energy of NMA by measuring the maximum desorption temperature. IRRA spectra were measured with a Bruker IFS 66/vs spectrometer equipped with a KBr beam splitter, CaF$_2$ windows, and an MCT detector. IRRA spectra were recorded with 4 cm$^{-1}$ resolution and 60 min scan time. All shown IRRA spectra were baseline corrected manually using the OPUS software's rubber band correction. Sample spectra were measured first, and the background spectra were measured next. The UHV setup for XPS was equipped with an XPS system consisting of a Specs Phoibos 150 electron energy analyzer and a 1D-DLD detector (Surface Concept 1D-DLD64_2−150) as well as a Specs Focus 500 monochromator including a Specs XR50 M X-ray source. All spectra were measured with monochromatic Al Kα radiation (1486.6 eV). Detailed XP spectra were measured with 100 ms dwell time, 0.05 eV step size, 10 eV pass energy, and 80 scans for carbon (C 1s), 20 scans for nitrogen (N 1s), and 2 scans for platinum (Pt 4f). All spectra were referenced to the bulk Pt 4f level at 71.1 eV.[28,29] Temperature-programmed XP spectra were measured by applying a heating ramp of 0.25 Ks$^{-1}$. TP-XP spectra were recorded with a 0.125 ms dwell time, 0.1 eV step size, and 30 eV pass energy.

## 3. RESULTS

In this section, we first present the results of the DFT calculations, followed by a summary of the relevant results from our experimental findings. The first task of DFT calculations is determining the adsorption site, which we carry out at the lowest coverage that we have considered. This is followed by a full analysis of the changes in the binding energy, the geometric and electronic structures, and the vibrational frequencies of the system as a function of NMA coverage.

**3.1. Theoretical Results.** *3.1.1. Determination of Adsorption Site of NMA on Pt(111).* Since NMA is a multiatom molecule with a nitrogen atom coordinating a phenyl ring and a methyl group, we followed a set of strategies to find the minimum energy configuration of NMA on Pt(111) at the low coverage of 1/36 (6 × 6 unit cell). This involved anchoring the nitrogen atom on the high-symmetry Pt(111) sites (top, bridge, and fcc and hcp hollow sites, shown in Figure S2), followed by ionic relaxation of the system. Once the nitrogen atom was found to bind at the top site, the phenyl ring was rotated by 30° to obtain the lowest energy configuration. We found three configurations in which the center of the phenyl ring was located at either the bridge, the fcc, or the hcp site, as shown in Figure 1a−c, respectively. Among the three configurations shown in Figure 1, Figure 1a represents the minimum energy configuration. The energy difference (ΔE) of the configurations in Figure 1b,c (phenyl ring center at the fcc site and hcp site with the nitrogen atom at the top site) relative to that in Figure 1a is 0.23 and 0.18 eV, respectively. The adsorption energy of NMA, $E_{ads}$, is calculated (eq 1 below) by subtracting the energy of the relaxed Pt(111) slab without the molecule ($E_{slab}$) and the energy of the relaxed NMA molecule in the gas phase ($E_{molecule}$) from the total energy of the system consisting of the adsorbed molecule on the Pt(111) surface after ionic relaxation ($E_{system}$).

$$E_{ads} = E_{system} - E_{slab} - E_{molecule} \quad (1)$$

Since the two functional groups in NMA, the phenyl ring and the amine, are similar structurally to benzene and ammonia, respectively, we have compared the adsorption characteristics of NMA with these two molecules (benzene and ammonia) on the Pt(111) surface, for which our calculated results (see SI-5) agree with those in the literature.[30,31] Like benzene,[30] the phenyl ring adsorbs at the bridge site on Pt(111), and as in the case of ammonia, the nitrogen of the amine group adsorbs at the top site of Pt(111),[31] as shown in Figure 1. In Section 3.1.2, we analyze the coverage-dependent adsorption behavior of NMA on Pt(111), using the configuration shown in Figure 1a as the reference. This reference site was consistently applied across all coverage levels without recalculating for each.

### 3.1.2. Coverage-Dependent Adsorption Characteristics of NMA on Pt(111).

#### 3.1.2.1. Adsorption Energy.

To examine the effect of coverage on NMA adsorption characteristics on Pt(111), we modeled the system using supercells of the following dimensions: 2 × 3, 3 × 3, 4 × 4, 5 × 5, and 6 × 6 unit cells. By definition of coverage as the ratio of the number of molecules to the number of surface atoms, the 2 × 3 unit cell with six surface atoms to one NMA (1/6 coverage) represents the highest coverage. The lowest coverage is the unit cell dimension 6 × 6, with 36 surface atoms (1/36 coverage). The coverage-dependent adsorption energy of NMA, including the zero-point energy correction, is summarized in Table 1. As is evident from Table 1, the NMA binding energy on the surface changes with coverage. It is −2.83 eV at the lowest coverage (1/36) and −1.43 eV at the highest (1/6). This decrease in binding energy with increasing coverage is expected since intermolecular interactions become increasingly important, causing the molecule to be less bound to the surface with increasing coverage.[32,33]

Table 1. Adsorption Energies for Each Coverage with and without the Zero-Point Energy (ZPE) Correction[a]

| coverage | adsorption energy $E_{ads}$ (eV) | ZPE contribution (eV) | $E_{ads} + E_{zpe}$ (eV) | Pt–N bond distances (Å) |
|---|---|---|---|---|
| 1/6  | −1.55 | 0.12 | −1.43 | 2.27 |
| 1/9  | −1.88 | 0.11 | −1.77 | 2.21 |
| 1/16 | −2.41 | 0.10 | −2.31 | 2.29 |
| 1/25 | −2.90 | 0.12 | −2.78 | 2.21 |
| 1/36 | −2.96 | 0.13 | −2.83 | 2.21 |

[a]The last column contains the bond distances between the surface and the nitrogen atom of NMA.

Table 1 shows that as the coverage decreases from 1/6 to 1/36, the Pt–N bond distance shortens from 2.27 to 2.21 Å, a reduction of 0.06 Å, along with a change in the molecule's binding strength. The Pt–N bond length is, however, not a good measure of the trend in the adsorption energy, as the molecule undergoes geometrical changes with coverage. As we discuss below, the phenyl ring tilts noticeably away from the surface at higher coverages because of the increased intermolecular interactions. This tilting correlates with the coverage-dependent displacement of the Pt and N atoms along the z-axis (see Figure S6(a,c)). The ensuing steering effects, with increasing coverage, force the phenyl ring to orient away from each other, resulting in a repulsive interaction that also affects the Pt−N bond length. Conversely, since the molecules are more dispersed at lower coverages, the reduced intermolecular interactions allow for stronger interactions between the carbon atoms of the phenyl ring and the platinum surface atoms than those at higher coverages. Consequently, the phenyl rings lie almost parallel to the surface at a low coverage. In short, molecular coverage plays a significant role in determining adsorption geometry and strength, with high coverage causing repulsion-induced tilting and the dominant molecule–surface interactions at low coverage making the molecule parallel to the surface.

#### 3.1.2.2. Geometrical Structure.

In Figure 2, we illustrate how the geometrical structure changes at different coverages through a top and side view of NMA adsorbed on Pt(111). At the high coverage (1/6), as seen in Figure 2a (ii), the carbon atoms of the phenyl ring lift off the surface and the molecule bonds to the Pt surface only through the nitrogen atom.

At an intermediate coverage (Figure 2b), NMA adsorbs via the nitrogen atom and five phenyl ring carbon atoms bind to three Pt atoms: two pairs of carbons ($C_2$, $C_4$ and $C_3$, $C_5$) share a Pt atom, $C_1$ binds to a third Pt atom, and $C_6$, connected to N, occupies a hollow site without direct contact with Pt. Since not all of the carbon atoms in the phenyl ring interact directly with the surface, the Pt–N bond length in this intermediate coverage experiences slight elongation compared to the low coverage. This is because the Pt atom is displaced downward toward the second layer of Pt (shown in Figure S6c) and the interaction between the nitrogen and the surface atom needs to compensate for the weaker interaction from the phenyl ring. This unique adsorption geometry causes a deviation in the Pt–N bond length at this intermediate coverage, which accounts for the anomalous value of 2.29 Å in Table 1 for 1/16 coverage. In Figure 2c, the top view of the adsorbed molecule at low coverage shows the NMA molecule to be in contact with five platinum surface atoms. Four Pt atoms interact with carbon atoms from the phenyl ring so that the two pairs of carbon atoms, ($C_5$,$C_6$) and ($C_1$,$C_2$), are each in contact with a Pt atom under them. Two single carbon atoms ($C_3$ and $C_4$) interact with two different Pt atoms under them, whereas the nitrogen atom interacts with the fifth Pt atom on the surface, as shown in the top view of Figure 2c (i). The methyl group carbon atom does not bind with the Pt surface, which is further discussed in Section 3.1.2 (iii).

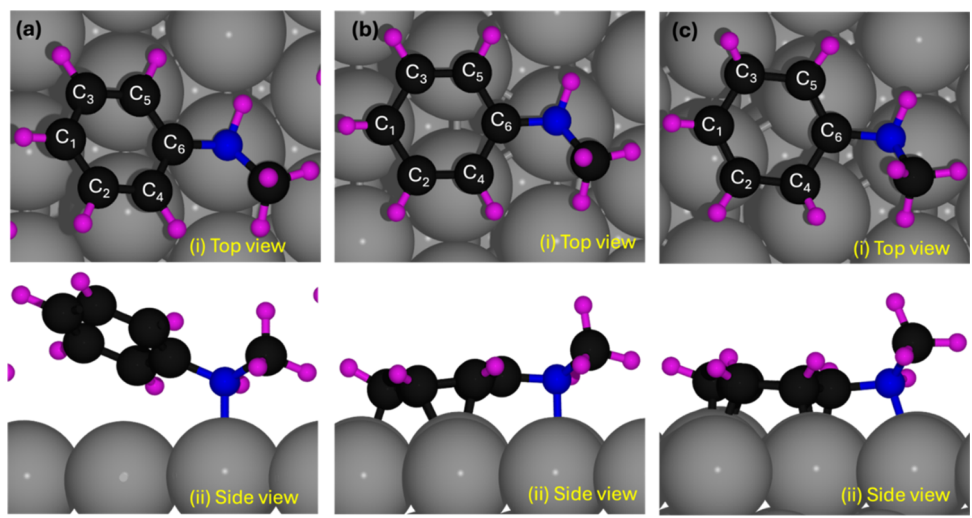

**Figure 2.** Top (i) and side (ii) views of the adsorbed molecule on (a) high coverage (1/6), (b) intermediate coverage (1/16), and (c) low coverage (1/36).

We also find that the C−C bond lengths of the phenyl ring carbon atoms show a slight increase of 0.1 Å for the low coverage (∼1.4 Å) as compared to that at the high coverage (∼1.3 Å), indicating changes in molecular conformation with increasing surface space. The bond length of carbon atoms of the phenyl ring for each coverage is compared to that of the gas phase NMA molecule in Table S4. There is also a decrease in the height of the nitrogen atom and the height of the center of the phenyl ring from the Pt(111) surface with decreasing coverage (see Figure S6a,b), indicating stronger interaction of the NMA with the surface at the lower coverage, as we have noted above. The tilt of the phenyl ring as a function of coverage is demonstrated in Figure 3. The inclination angle decreases from 33.5° to 2.8° as the coverage changes from 1/6 to 1/36. Figure 3 also captures the relationship between the phenyl ring's inclination angle and the NMA molecule's adsorption energy trend with respect to coverage on the Pt(111) surface. At 1/36, the molecule adsorbs strongly on Pt(111) through the phenyl ring and the nitrogen atom, and the adsorption strength decreases at 1/6, for which the adsorption occurs primarily through the nitrogen atom (numerically shown in Table 1).

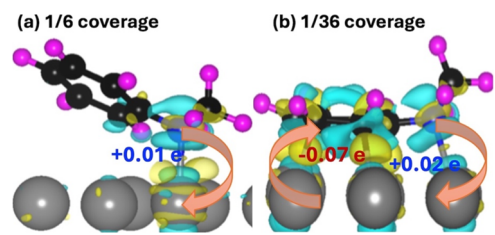

Figure 4. Charge density difference plot for (a) 1/6 coverage and (b) 1/36 coverage showing the charge accumulation (yellow) between the nitrogen atom and Pt atom on the surface.

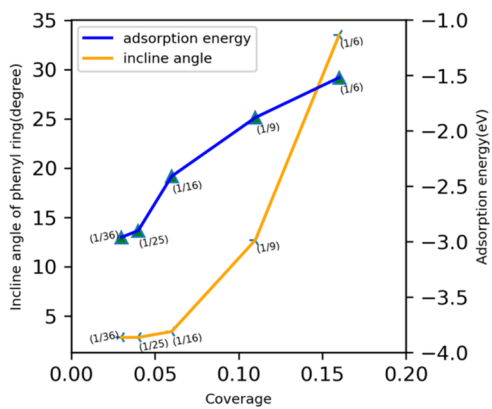

Figure 3. Coverage dependence of the inclination of the phenyl ring (orange) and NMA adsorption energy (blue).

*3.1.2.3. Electronic Structure of NMA on Pt(111).* For a key understanding of the coverage-dependent chemical activity of NMA on Pt(111), we turn to insights into the electronic structure of the system as obtained through a detailed analysis of charge transfer (through the Bader charge in Tables S1 and S2), charge density distribution (Figure 4), and projected electronic density of states, as shown in Figure 5. In Figure 4a, the charge density difference (eq S2) was used to calculate the charge redistribution of the 1/6 system, which shows that charge sharing occurs between the nitrogen atom and the surface Pt atom, with the former losing charge to the later (+0.01 e). As the molecule lifts away from the surface, facilitated by the phenyl ring at a high coverage (Figure 4a), no charge transfer occurs between the carbon atoms of the phenyl ring and the surface. The molecule interacts with the Pt surface solely through the nitrogen atom (with a bond length of 2.27 Å). These findings are in accordance with our experiments that demonstrate that the molecule detaches intact from the surface at monolayer coverage as the temperature increases, as discussed in Section 3.2. This indicates a strong tendency for NMA to separate from the surface at high coverage by breaking the bond between the nitrogen atom and the surface, as illustrated in Figure 4a. Conversely, at low coverage, NMA exhibits a stronger interaction with the surface (as shown in Table 1) through both the phenyl ring and the nitrogen atom.

Bader charge analysis (Table S2) shows that the carbon atoms of the phenyl ring gain charge from the Pt surface as the sum of the charge on each carbon atom of the phenyl ring is −0.07e, and the nitrogen atom looses 0.02e. Correspondingly, the charge difference plot in Figure 4b (from eq S3) displays a small charge sharing between the molecule and the surface. At low coverage, charge transfer from the phenyl ring's π-system to the metal reduces the ring's π-character, as indicated by the charge redistribution in Figure 4b and also by the elongation of C−C bonds, as shown in Table S4 in the SI. Note that while the charge transfer between the molecule and the surface is small, almost within the error in DFT, there is a striking coverage-dependent charge redistribution at the interface as depicted in Figure 4.

The projected electronic density of states plotted for the molecule (Figure S4), the adsorbed system (Figures S7−S11), and the pristine slab (Figure S3) provide further insights into the bonding characteristics of NMA on Pt(111). In the case of the high coverage (Figure 5a), there is no hybridization among the phenyl ring's carbon atom and the surface Pt atoms, but the $p_z$ orbital of the nitrogen atom hybridizes with Pt $d_{z^2}$ orbital. After the adsorption of the NMA at 1/36 coverage, the peaks of the density of states of the adsorbed system shift to lower energies compared to the clean Pt(111) surface (Figure S3), representing a sharing of electrons among the surface and the molecule which is also seen in the charge difference plot Figure 4b. The projected density of states (Figure 5b) for the atoms of the molecule that are in contact with the Pt surface atoms show that the $p_z$ orbital of the nitrogen atom and those of the carbon atoms of the phenyl ring hybridize with Pt $d_{z^2}$ orbital. The above reasoning leads to the conclusion that the molecule's propensity to dissociate into fragments is increased at low coverages, while that to desorbs intact prevails at higher coverages as the molecule is not strongly bound to the Pt surface.

*3.1.2.4. Vibrational Frequencies of NMA on Adsorption on Pt(111).* Changes in the vibrational frequencies of molecules from their gas phase values on adsorption on surfaces provide crucial information about molecule−surface interactions and system structural dynamics. This is particularly true of the symmetric and asymmetric stretching and in-plane and out-of-plane bending vibrations of the molecules. Below we turn to an analysis of the vibrational modes of NMA in the gas phase, which is followed by that of NMA adsorbed on Pt(111) for the coverages considered in this work.

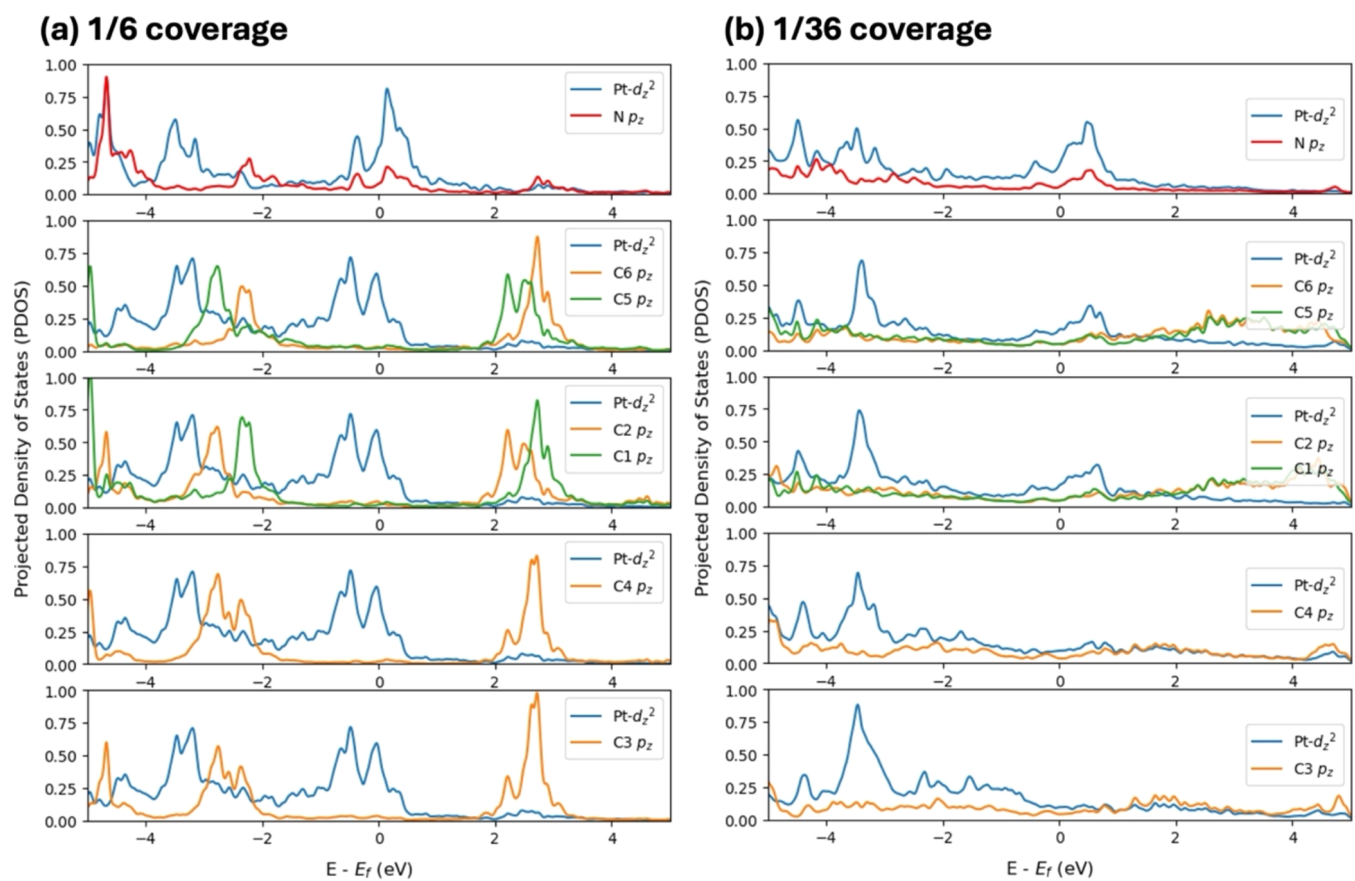

**Figure 5.** PDOS of (a) 1/6 coverage where the nitrogen pz orbital hybridizes with the Pt $d_{z^2}$ orbital but the carbon p-orbitals of the phenyl ring do not hybridize with the Pt $d_{z^2}$ orbital, and that of (b) 1/36 coverage where the carbon pz orbitals (from C1 to C6) of the phenyl ring and the nitrogen pz orbital hybridize with the Pt $d_{z^2}$ orbital below the Fermi energy.

**Table 2. Summary of Experimental and Calculated Vibrational Frequencies ($cm^{-1}$) of NMA on Pt(111) Presented in This Work and Those Reported in the Literature**

| 3.5 ML NMA (experiment) | 1.0 ML NMA (experiment) | 0.3 ML NMA (experiment) | assignment | literature[11,39] | 1/6 coverage NMA (theory) | 1/36 coverage NMA (theory) |
|---|---|---|---|---|---|---|
| 2880 | 2882 | 2880 | $\nu(CH_3)_{asym.}$ | 2880 | 3027.4 | 3046, 3058, 3065.8 |
| 2813 | 2815 | 2812 | $\nu(CH_3)_{sym.}$ | 2811 | 2942.7 | 2971.7 |
| 1608 | 1606 | 1605 | $\nu(C-C)_{ring}$ | 1608 | 1584 | |
| 1586 | 1586 | | $\nu(C-C)_{ring}$ | | 1580 | |
| 1508 | 1504 | 1504 | $\nu(C-C)_{ring}$ | 1509 | | |
| 1477 | 1476 | | $\nu(C-C)_{ring}$ | | 1475.1 | 1413.9 |
| | 1445 | 1450 | $\delta(CH_3)_{asym}$ | | 1445, 1418, 1399 | 1447.3, 1428.9 |
| 1339 | 1339 | | $\nu(C-N)$ | | 1295, 1350 | |
| 1324 | | | $\nu(C-C)_{ring}$ | 1324 | | 1350 |
| 1265 | 1261 | | $\delta(CH)_{in-plane}$ | 1268 | | 1252 |
| 1181 | 1179 | | $\delta(CH)_{in-plane}$ | 1181 | 1216 | |
| 1152 | 1152 | | $\delta(CH)_{in-plane}$ | 1152 | 1162 | 1143 |
| 1127 | | | $\delta(CH)_{in-plane}$ | 1128 | 1120 | 1115 |
| 1072 | 1072 | 1072 | $\delta(NH)_{out\ of\ plane}$ | 1074 | | |

*3.1.2.5. Gas Phase NMA Molecule.* In our calculations of vibrational frequencies for NMA in the gas phase, the in-plane C−H bending modes $\delta(CH)_{in-plane}$ of the phenyl ring (Figure S13) and the $CH_3$ group $\delta(CH_3)_{sym}$ attached to the amine (Figure S15) are found within the 1000−1500 $cm^{-1}$ range, consistent with reported experimental values.[11,34] The in-plane N−H bending mode $\delta(N-H)_{in-plane}$ appears at 1525.8 $cm^{-1}$, while the symmetric and asymmetric $CH_3$ stretching modes are observed at ∼2900−2980 $cm^{-1}$ $\nu(CH_3)_{sym}$ (Figure S18) and ∼3000−3030 $cm^{-1}$ $\nu(CH_3)_{asym}$ (Figure S19), respectively. Additionally, the high-frequency N−H stretching of the amine group is found at 3540.50 $cm^{-1}$ $\nu(N-H)_{sym}$ (Figure S20), which also aligns well with values reported for these segments of NMA.[11,34] Since the NMA molecule is versatile in its geometrical structure, with the phenyl ring and methyl group connected to the amine group, we have resorted to comparisons of vibrational frequencies calculated of its components for validation of those we obtain here. Comparison with experimental results has relied on data for NMA molecules in the liquid phase, as that is what is available (Table 2). Differences between the experimental and theoretical values can occur, particularly for the oscillations at higher wavenumbers, since in this range the DFT method tends to overestimate the frequencies. The values given in this work were not corrected using a scaling factor.

*3.1.2.6. NMA Adsorbed on Pt(111).* Figure 6 presents the simulated infrared (IR) spectra for the NMA molecule in the gas phase and adsorbed on the Pt(111) surface at each considered coverage. The vibrational modes have been assigned by analyzing the displacement vector directions depicted in the patterns (Figures S13−S20). These assignments were compared with data from the literature and experimental results to ensure accuracy and validity (Table 2). The vibrational frequencies of the platinum modes predominantly fall within the low-frequency range, as shown in Figure 6 (top row). It is important to note that the most significant contributions to the observed spectra in the high-frequency range are attributed to molecular adsorption on the metal surface, highlighting the substantial influence of the adsorbed species on the spectral features.

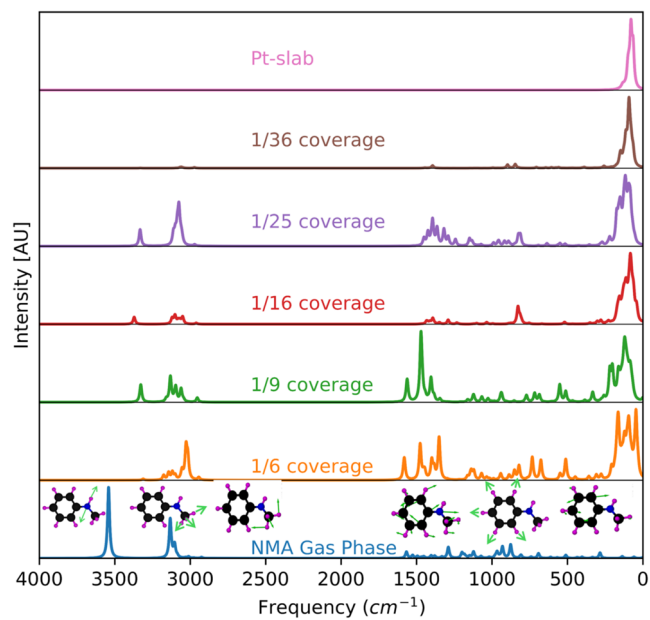

**Figure 6.** Simulated IR spectra of isolated and adsorbed NMA molecules as a function of coverage on the Pt(111) surface. Coverage ranges from 1/36 (low coverage) to 1/6 (high coverage). The first row (top) represents the vibrational modes of the clean Pt slab, while the last row (bottom) shows the vibrational modes of the molecule in the gas phase. Displacement patterns for the gas phase vibrational modes are shown with the corresponding frequency range, and detailed visualizations are provided in the Supporting Information (SI).

The analysis of the mode frequencies reveals a general trend of decreased vibrational frequencies upon adsorption, indicating weakening of the molecular bonds. For instance, the N−H stretch mode of the amine group exhibits a reduction in frequency compared with that in the gas phase molecule, signifying bond weakening and redistribution of electron density upon adsorption in both low and high coverage. Similarly, the C−H in-plane deformation mode and the C−C ring stretch vibrational mode also display decreases in frequency upon adsorption, suggesting a weakening of the respective bonds due to interactions of NMA with the Pt(111) surface.

On the contrary, the $CH_3$ symmetric stretch mode shows an opposite trend: it increases upon adsorption. Note that the methyl group is not directly bonded to the Pt surface but is indirectly connected through the nitrogen atom, which adsorbs on the Pt surface. The above trend highlights how the vibrational response of specific functional groups can differ depending on the nature of their interaction with the substrate and with other entities in the molecule. Of course, in isolating the contributions to the vibrational modes of a multi-component, chemisorbed molecule we need to be aware that this is a limited view as the constituent groups are all connected.

*3.1.3. Decomposition of NMA on Pt(111).* The theoretical analysis outlined above suggests the potential decomposition of NMA on Pt(111) at low coverages. This is confirmed by experimental TPD and XPS data presented below, which indicate the temperature-dependent decomposition of NMA. To gain some mechanistic insights, we present here a summary of our findings on the thermodynamics of a few chemical reactions of interest at two coverages. A full analysis of all reaction pathways is beyond the scope of this work that focuses on understanding the coverage-dependent adsorption characteristics of NMA on Pt(111).

Our study of reaction energies evaluates the activation of the different bonds in NMA with the aim of discovering whether there is a preference for the activation of the C−N, N−H, or C−H bonds of the methyl or phenyl groups in NMA. Calculations were conducted to determine the bond dissociation energy by breaking the bond of interest and comparing the energy to that of the adsorbed NMA molecule for coverages of 1/36 and 1/9.

At 1/36 coverage of NMA on Pt(111), breaking of the C−N bond between the nitrogen atom and the carbon atoms of the phenyl ring was found to be an endothermic reaction, requiring 1.43 eV, as indicated in Figure 7a. Similarly, Figure 7b illustrates the ionic relaxation configuration of the benzene ring and $NCH_3$ moiety resulting from dissociating a hydrogen atom from the amine, with a reaction energy of 0.09 eV. Additionally, Figure 7c shows the dissociation energy of the methyl group from NMA on the Pt(111) surface, which was determined to be 0.21 eV. Furthermore, Figure 7d presents the dissociation energy of the hydrogen atom connected to the nitrogen atom, which is also an endothermic reaction with an energy increase of 0.59 eV. Finally, Figure 7e,f represents the dissociation energies of one and two hydrogen atoms from the methyl group carbon atom, which were found to be exothermic, releasing energies of −0.22 and −0.31 eV, respectively.

These results suggest that the activation of the methyl C−H bond is energetically favored compared to other bonds in NMA. Activating the first C−H bond is an exothermic step that lowers the energy by −0.22 eV, while activating the second one reduces the energy further by −0.09 eV. From here, the formation of hydrogen cyanide may occur by breaking the nitrogen phenyl bond and transferring the hydrogen bonded to the nitrogen atom to the phenyl ring, forming benzene and hydrogen cyanide, as seen in Figure 7b. Another potential pathway for the decomposition of the NMA molecule is the replacement of the nitrogen phenyl bond with hydrogen, producing benzene and methylamine molecules on the surface. The methylamine molecule may be further decomposed to form hydrogen cyanide.

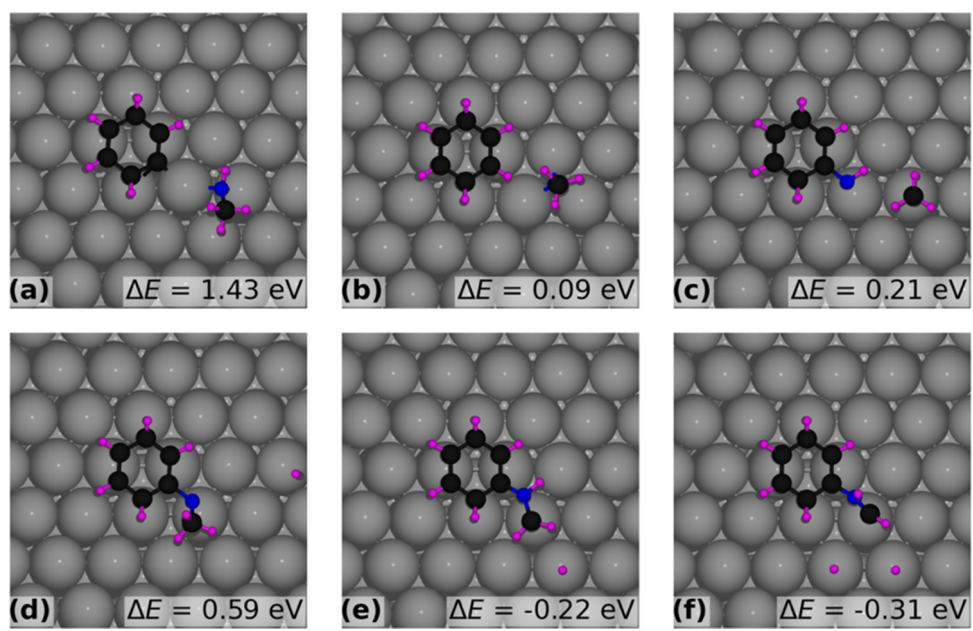

Figure 7. Reaction energy ΔE (difference in total energy between the illustrated structure and that with the intact NMA molecule on Pt(111)) for fragment formation at 1/36 coverage of NMA on Pt(111): (a) $C_6H_5$ and $CH_4N$; (b) $C_6H_6$ and $CH_3N$; (c) $C_6H_6N$ and $CH_3$; (d) $C_7H_8N$ and hydrogen atom dissociated from nitrogen; (e) $C_7H_8N$ and the hydrogen atom dissociated from the methyl group; and (f) $C_7H_7N$ and two hydrogen atoms dissociated from the methyl group.

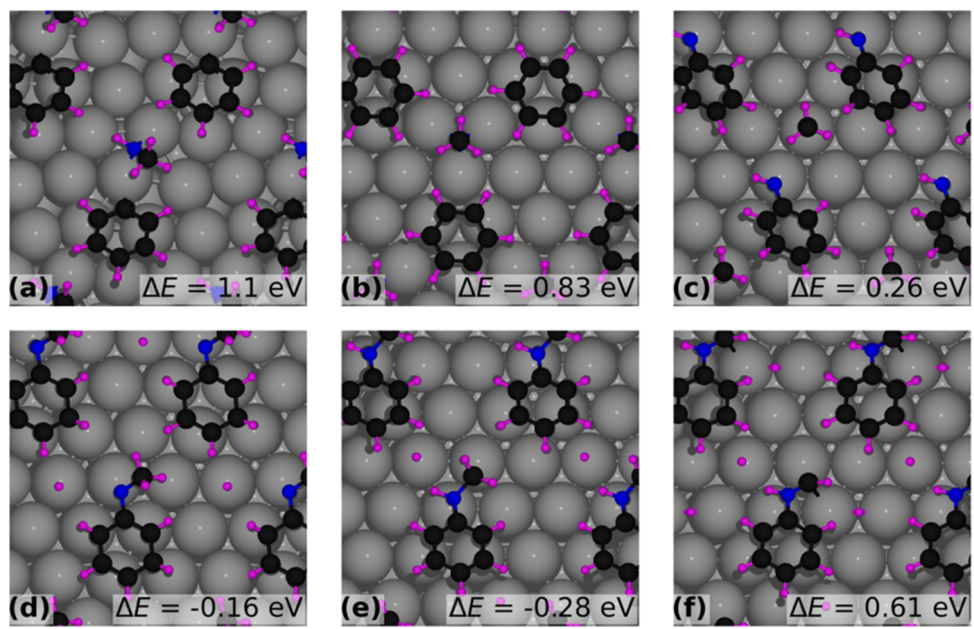

Figure 8. Reaction energy ΔE for fragment formation at 1/9 coverage of NMA on Pt(111): (a) $C_6H_5$ and $CH_4N$; (b) $C_6H_6$ and $CH_3N$; (c) $C_6H_6N$ and $CH_3$; (d) $C_7H_8N$ and hydrogen atom dissociated from nitrogen; (e) $C_7H_8N$ and the hydrogen atom dissociated from the methyl group; and (f) $C_7H_7N$ and two hydrogen atoms dissociated from the methyl group.

To understand the effect of NMA coverage on the activation of the bonds, the reaction energy for breaking each bond shown in Figure 7 for the low coverage may be compared with those in Figure 8 for the higher coverage of 1/9. It can be seen from Figures 7a and 8a that breaking the C−N bond with the phenyl ring is energetically costly for both coverages, with reaction energies of 1.43 and 1.1 eV, respectively. Conversely, breaking this bond is much more favored if the carbon is bonded with hydrogen, particularly at low NMA coverage, as indicated by Figures 7b and 8b, showing reaction energy at low and high coverages of 0.09 and 0.83 eV, respectively. Furthermore, breaking the C−N bond with the methyl group yields about the same energy for each system: 0.21 eV for low coverage and 0.26 eV for high coverage (Figures 7c and 8c). The results for breaking the N−H bond shown in Figures 7d and 8d yield varying results for the coverage; for low coverage, this bond breaking is unstable, increasing the energy by 0.59 eV, while for high coverage, this results in lowering the energy by −0.16 eV. For the low coverage, it can be seen in Figure 7e,f that the methyl group can be easily dehydrogenated 2 times, first lowering the energy by −0.22 eV and then further reducing the energy down to −0.31 eV, laying the potential pathway for the decomposition of the NMA molecule. Dehydrogenation of the molecule in the high-coverage regime is not preferred as the first step lowers the energy to −0.28 eV, and the second step is less favored. In the second step, as shown in Figure 8f, removing the second hydrogen results in more energy, increasing to 0.61 eV. The above results provide a qualitative understanding of the propensity for NMA to decompose at lower coverages and desorb intact at the higher coverage considered here.

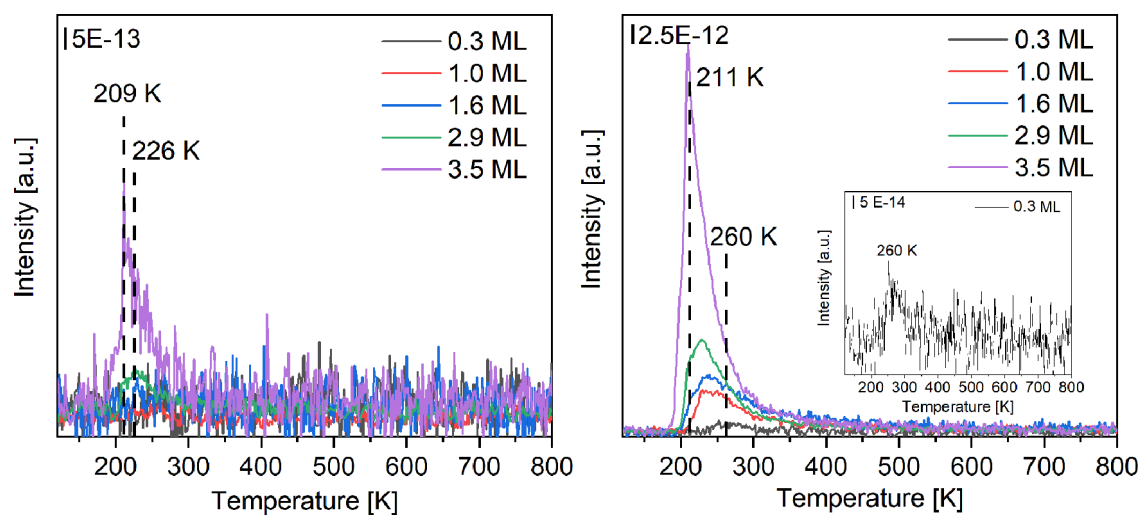

Figure 9. Coverage-dependent temperature-programmed desorption spectra (TPD) of adsorbed *N*-methylaniline at Pt(111) for mass fragments $m/z$ = 78 (left) and 106 (right). *N*-Methylaniline was dosed through a pinhole doser filled with 1 mbar, while the crystal was cooled down ($T$ < 110 K). The inlet shows the TPD data of 0.3 ML NMA.

**3.2. Experimental Results. 3.2.1. Coverage-Dependent Temperature-Programmed Desorption Spectra (TPD) of N-Methylaniline on Pt(111).** To support the findings from the theoretical calculations and to gain further insight into the surface chemistry of NMA, the coverage-dependent adsorption of NMA on Pt(111) was investigated using surface-sensitive methods such as X-ray photoelectron spectroscopy (XPS), temperature-dependent desorption (TPD) and Fourier transform infrared reflection absorption spectroscopy (FT-IRRAS). Figure 9 shows TPD spectra for the mass fragment $m/z$ = 106 and 78 collected after varying the dosing time and thus the coverage. The coverages are given in multiples of a monolayer, which is the experimentally determined maximal possible packing of the molecules in direct contact with the surface, using the integrals of the TPD spectra and assuming that the red curve is related to a coverage close to a saturation coverage.

The fragments of *N*-methylaniline and its decomposition products were identified by references from the NIST database[35,36] and literature.[37] Table S5 shows the contributors to different mass fragments, and Figure S23 shows all of the measured mass fragments for an adsorbed multilayer of NMA on Pt(111). For calculation of the desorption energy, see Figure S24. The mass fragment $m/z$ = 78 (benzene) in Figure 9 shows a very weak desorption peak at 260 K for 0.3 ML coverage, the signal intensity of which is just above the ground noise of the spectrum. On increasing the coverage to 1.0 ML, the signal shifts to 230 K. The development of a shoulder at 211 K is apparent in the blue curve. The additional desorption feature at 211 K does not reach saturation with increasing coverage and can be assigned to multilayer desorption. The 230 K feature in the blue curve is still slightly larger than in the red curve according to curve subtraction, although the percentage of growth is small. Therefore, calibrating the TPD data by setting the red curve to the equivalent of a monolayer may slightly overestimate the coverage. Simultaneous growth of the monolayer and multilayer indicates that the adsorption behavior is more of a Stranski-Krastanov growth. After reaching the monolayer, a sudden increase in intensity of the multilayer signal is apparent though the dosing times have not increased significantly. The dosing time via the pinhole doser between the purple curve and the blue curve differs by a factor of 1.5, while the peak integrals differ by a factor of 2.8. This indicates a change of sticking probabilities after reaching a monolayer. The monolayer desorption peak of NMA at 230 K and the multilayer desorption at 211 K observed here are in good accord with aniline monolayer desorption at 240 K and multilayer desorption at 200 K on Pt (111).[11]

Interestingly, the molecular desorption of NMA ($m/z$ = 106) can be observed only after the occurrence of the multilayer desorption peak in the TPD spectrum of $m/z$ = 78 at 210 K at 1.6 ML coverage. At lower coverages, however, lower mass fragments are observed which may be attributed to dissociation of NMA. Besides the benzene fragment, the following fragments were observed (see Figure S23): $H_2$ ($m/z$ = 2), HCN ($m/z$ = 27), $CO/N_2$ ($m/z$ = 28), $MeNH_2$ ($m/z$ = 30), and $C_2N_2$ ($m/z$ = 52).

**3.2.2. Coverage-Dependent FT-IRRAS Spectra of N-Methylaniline on Pt(111).** Before performing FT-IRRAS measurements, TPD spectra were measured to identify the sub-monolayer, monolayer, and multilayer coverage. Since we used $CaF_2$ windows, which are impermeable to infrared light below 1000 cm$^{-1}$, no spectra could be collected at lower wavenumbers. Note that the absorption bands of water from the atmosphere clogged the region above 3000 cm$^{-1}$. Even though the sample room of the spectrometer was flushed with dry compressed air, small amounts of water were still present. Unfortunately, both regions exhibit some analytically useful absorption bands. The N−H stretch vibration or the C−H out-of-plane bending mode was not considered under the described experimental conditions. Since the spectral region between 2.700 and 1.650 cm$^{-1}$ did not exhibit any analytically interesting band, it was cut out of the spectrum. Figure 10 shows the coverage-dependent FT-IRRAS spectra of N-methylaniline on Pt(111), beginning from the top with the sub-monolayer coverage (0.3 ML) (blue), 1.0 ML coverage (red), and with the multilayer coverage (3.5 ML) (black).

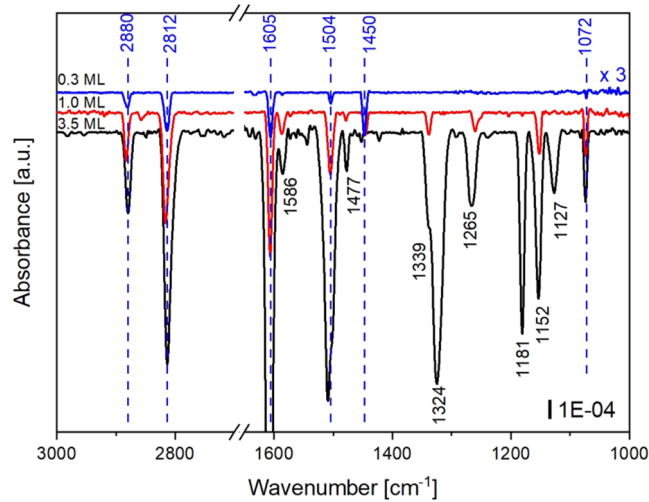

Figure 10. Coverage-dependent FT-IRRAS spectra of N-methylaniline on Pt(111) measured after dosing a sub-monolayer (0.3 ML), a monolayer (1.0 ML), and a multilayer (3.5 ML) of N-methylaniline at liquid nitrogen temperature through a pinhole doser filled with 1 mbar. The sub-monolayer spectrum was scaled by a factor of 3 for better visibility. Wavenumbers marked in blue were present at sub-monolayer coverage, while wavenumbers marked in black evolved with increasing coverage. An overview of the calculated and experimentally observed absorption bands and the assignment of the absorption bands in prior work[11,34,38,39] are given in Table 2.

At sub-monolayer coverage (0.3 ML), six absorption bands at 2880, 2812, 1605, 1504, 1450, and 1072 cm$^{-1}$ can be observed experimentally. The two absorption bands at 2880 and 2813 cm$^{-1}$ can be assigned to the asymmetric $\nu(CH_3)_{asym}$ and symmetric stretch vibration $\nu(CH_3)_{sym}$ of the methyl group and are in good accordance with prior results obtained for NMA on Pt (111) and liquid NMA.[11,34,39] The absorption bands at 1605 and 1504 cm$^{-1}$ relate to the ring vibration $\nu(C-C)_{ring}$ and the absorption band at 1072 cm$^{-1}$ to the $\delta(NH)_{out,of,plane}$ mode.[11,34]

On increasing the coverage to 1.0 ML (red curve), the absorption bands for the ring vibration $\nu(C-C)_{ring}$ at 1605 and 1504 cm$^{-1}$ exhibit a significant increase in intensity by almost a factor 4, while the deformation vibration of the methyl group $\delta(CH_3)_{asym.}$ at 1450 cm$^{-1}$ decreases. Further absorption bands at 1586 cm$^{-1}$ ($\nu(C-C)_{ring}$), 1339 cm$^{-1}$ ($\nu(C-N)$), 1265 cm$^{-1}$ ($\delta(CH)_{in-plane}$), 1181 cm$^{-1}$ ($\delta(CH)_{in-plane}$), and 1152 cm$^{-1}$ ($\delta(CH)_{in-plane}$) appear to indicate an up-tilting of the ring. Due to the metal surface selection rules, a transition dipole moment parallel to the metal surface interferes destructively with the induced image dipole within the metal.[40,41] This is why these components of the vibrational modes are not apparent in the spectra. Therefore, the intensity changes indicate an almost parallel orientation of the ring to the surface at low coverages and an up-tilting of the ring with increasing coverage. This coverage dependency is in accordance with the results from our theoretical calculations summarized in Figure 6. The coverage-dependent changes underline the results of the theoretical calculations of a significant coverage-dependent change in the angle between the ring and the surface.

For multilayer coverage, the spectra exhibit two additional bands at 1324 and 1127 cm$^{-1}$, which may be assigned, respectively, to a $\nu(C-C)_{ring}$ mode and $\delta(CH)_{in-plane}$ mode. The appearance of these two modes may be attributed to random orientations of NMA on the surface within the multilayer. The observation is in good accord with results from the literature on a multilayer of NMA at Pt(111), liquid NMA as well as aniline derivates.[11,34,38,39,42,43]

We also note some differences in our results compared to those in the literature. The N−H bending mode, which is reported for o- and m-chloroaniline in the liquid phase at 1617 cm$^{-1}$,[42] for liquid aniline at 1618 cm$^{-1}$,[43] and for liquid NMA at 1620 cm$^{-1}$,[39] was not observed in the experimental spectra in this study. This is in accordance with the results of Trenary and co-workers on NMA on Pt(111).[11] Finally, note that the results of the calculated symmetric and asymmetric stretching vibrations of the methyl group in Table 2 deviate from the experimental values because of overestimation in the theoretical calculations of the high-frequency vibrations. Note that we have not applied the oft-used scaling factor when comparing experimental and theoretical results.

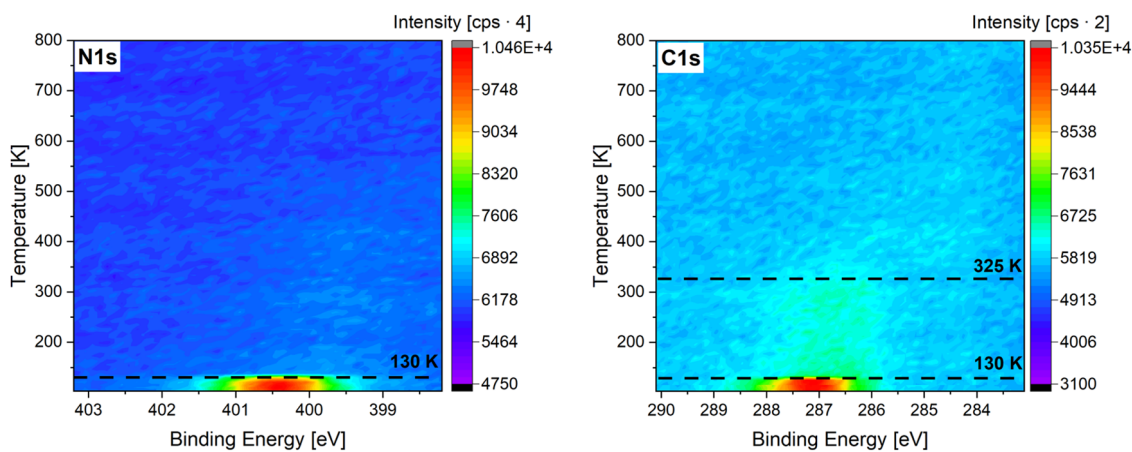

Figure 11. Temperature-programmed XP spectra of an adsorbed multilayer of N-methylaniline (NMA) at 110 K on Pt(111) with the N 1s region (left) and the C 1s region (right) in the temperature range 120−800 K.

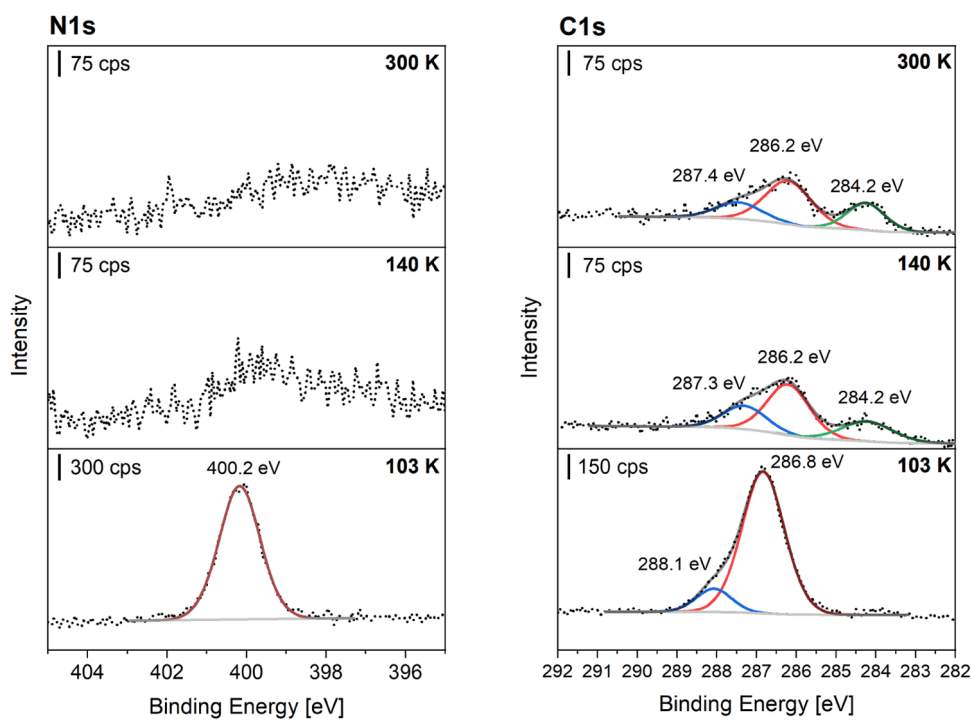

**Figure 12.** XP spectra of an adsorbed multilayer of *N*-methylaniline (NMA) at 110 K on Pt(111) followed by heating to the specified temperatures with the N 1s spectra (left) and the C 1s spectra (right). The Pt(111) single crystal was heated for 2 min to elevated temperatures and cooled down afterward.

*3.2.3. Temperature-Programmed X-ray Photoelectron Spectra of a Multilayer N-Methylaniline Adsorbed on Pt(111).* Temperature-programmed XPS (TP-XPS) allows us to follow the desorption or dissociation of an adsorbed molecule on the surface as a function of temperature. Figure 11 shows TP-XP spectra of a multilayer *N*-methylaniline adsorbed on Pt(111) at 110 K.

The temperature-dependent X-ray photoelectron (TP-XP) spectra in Figure 11 show a strong decrease in signal intensity at 130 K in the C 1s spectra and no signal above 130 K in the N 1s spectra. The TP-XP spectra indicate that the multilayer of NMA desorbs molecularly, while a small amount of NMA partially dissociates to carbon residues that remain at temperatures up to 325 K at the platinum surface. The remaining species desorb at 325 K, as no further signal is observed at higher temperatures. For gaining further insight into the adsorbed species, a multilayer of NMA was adsorbed on Pt(111) and detailed C 1s and N 1s spectra with a higher resolution were measured after heating to 140 and 300 K. Figure 12 shows the detailed C 1s and N 1s spectra at elevated temperatures.

After the adsorption of NMA at 103 K on the Pt(111) surface, the N 1s spectrum exhibits one intense peak at 400.2 eV, corresponding to the amine group of NMA. The measured N 1s binding energy is in a typical range for an amine functional group (N–H) on platinum surfaces.[4,5,17,44] The C 1s spectrum shows two features at 286.8 and 288.1 eV with an intensity ratio of 6:1. The peak at 288.1 eV can be attributed to the carbon atom of the methyl group and the peak at 286.8 eV to the six carbon atoms of the phenyl ring of NMA. However, the values obtained in this work for the C 1s species at 103 K, are up-shifted by 1.6 to 1.9 eV compared to literature results for NMA on Pt(111) at 300 K.[11] Oxygen contamination can be ruled out as a possible reason for an upshift (see Figure S25). The Bader charge analysis (Tables S1, S2) and Figure 4a have shown that electron density is donated from the nitrogen atom to the metal surface at high coverage. The charge transfer would result in the nitrogen atom being partially positively charged, which could be compensated by charge transfer from the ring by the mesomeric effect. The C 1s species measured in this work are comparable to the measured binding energies for carbon atoms adjacent to NH+ groups (287.9 to 287.6 eV) for polyaniline films[45,46] However, the delocalization of the positive charge could not be observed in polyaniline films because of possible hindrance[37] and the N 1s signal lay in a typical range for amines on the Pt surface and no significant upshift was observed for the amine group because of a lack of charge. Therefore, this strong upshift is according to the charging effects of the multilayer.

Comparing the 300 K spectra with the literature results, there is still an upshift of 0.9 to 1.2 eV of the C 1s signals. This difference compared to the literature for the C 1s binding energies is due to thermal changes that occur when heating above 140 K, as detailed N 1s spectra do not show any amine feature above 140 K while the C 1s features are downshifted by 0.6 and 0.8 eV compared to the spectra after adsorption of NMA at 103 K. Furthermore, the peak area ratio between both C 1s species change from 6:1 after adsorption to approximately 2:1 after heating to 140 K and a new species at 284.2 eV appears. Further heating to 300 K leads to no further changes.

The thermal changes in the XP spectra can be attributed to the formation of a new surface species by dehydrogenation or dissociation of NMA. The remaining carbon species above 140 K is probably due to the phenyl ring since methane and methyl groups have been observed at lower binding energies between 282.6 and 283.5 eV in the literature.[47,48] The evolution of a shoulder at lower binding energies at around 284 eV has also been observed for toluene and methylcyclohexane at Pt(111) and has been attributed to the formation of benzyl species by C−H scission.[49] Since NMA shows a similar behavior compared to toluene, a partial dehydrogenation of NMA is likely. However, a concrete assignment of the species at 284.2 eV is still difficult since it would fit both C−H fragments and carbonization of the platinum surface by decomposition of NMA.[48,50,51] Therefore, this peak is assigned to a species resulting from the coking of the surface with $C_xH_y$ fragments.

## 4. DISCUSSION

The aim of this study was to elucidate the adsorption and surface chemistry of *N*-methylaniline on Pt(111). NMA served as an appropriate model molecule because of its diverse functionalities. It is an aromatic secondary amine and exhibits N−H, C−N, aromatic C−H, and aliphatic C−H bonds, which makes it possible to check which of the bonds is activated first. Below we discuss the investigated coverage-dependent thermal changes to highlight possible essential steps in the activation of aromatic amines that provide the basis for potential heterogeneously catalyzed applications.

**4.1. Coverage-Dependent Amine−Surface Interaction.** At low coverage, NMA prefers to adsorb in the bridge orientation with its phenyl ring parallel to the Pt(111) surface, enhancing bonding that includes partial charge transfer from the platinum to the ring. The C−C ring bond length is increased, which indicates a weakening of the double bond character. However, with increasing coverage, the interaction of NMA with the platinum surface decreases as the aromatic ring becomes up-tilting from the surface because of lateral repulsive interactions. This up-tilting of the aromatic ring leads to an increase in the Pt-ring and Pt-nitrogen distance. The decrease in NMA adsorption energy by almost a factor of 2 for the coverages considered here attests to the important role of the aromatic ring in the interactions of NMA with the platinum surface.

The strong interaction of the NMA with the metal surface at low coverages (through two entities) makes its decomposition likely, while at higher coverages the increased Pt−N and Pt-ring distances concomitant with weaker interaction favor molecular desorption. The strong interactions that ensue when the aromatic ring lies parallel to the metal surface may cause a destabilizing effect on the amine group. The decreasing distance between the metal surface and the C−N bond between the ring and amino group may yield an easier cleavage of this bond, thus promoting desorption of the amine group. This effect on the activation of the C−N bond by parallel adsorption of the aromatic ring has been observed for the hydrogenolysis of aniline and for the adsorption of (*S*)-(−)-1-(1-naphthyl)ethylamine at Pt(111).[16,17] A parallel configuration of the adsorbed aniline or (*S*)-(−)-1-(1-naphthyl)-ethylamine ensures a strong interaction with the metal surface and facilitates the hydrogenolysis or secession of the ethylamine moiety, respectively.[16,17]

**4.2. Activation of C−N Bond versus N−H and C−H Bonds.** In the literature, we find that the activation of the C− N bond is favored in the parallel orientation of the ring of the aromatic amine. In contrast to the results mentioned above for aromatic amines, aliphatic amines show that the dehydrogenation via cleavage of the N−H and C−H bonds is preferred over the activation of the C−N bond.[6,7,10,52] Methylamines tend to form an aminocarbyne species through dehydrogenation, which further decomposes to form cyanide, hydrogen cyanide, or methyl-cyanide at the platinum surface.[6,7,10,52] The reason for the easier activation of the N−H and C−H bonds for methyl- and dimethylamine lies, on the one hand, in the stabilization of the intermediates through the formation of C− N double and triple bonds, which are stabilized by π-interactions with the surface. On the other hand, the strength of the different bonds plays a role.[52−54]

The question arises whether a phenylaminocarbyne species can form as an intermediate on the platinum surface and whether the cleavage of the C−N bond is carried out in a subsequent step. One might expect the formation of this species from the surface chemistry of the methylamines. To find an answer to this question, theoretical calculations were carried out to identify the elementary steps toward hydrogen cyanide formation. The cleavage of the C−H bond is found to be preferred over other bonds since the nitrogen atom is more stable when the C−H bond is activated than when the N−H bond is activated. The decomposition of NMA starts with dehydrogenation of the C−H bonds, leading to the first step in the formation of a phenylaminocarbyne species, which is in accordance with the literature on methylamines. Furthermore, the cleavage of the C−N bond between the aromatic ring and the amine group occurs after further activation of the C−H and N−H bonds. Moreover, the cleavage of the C−N bond is a consequence of previous bond activations and leads to the elimination of hydrogen cyanide. From the above results, we conclude that the aminocarbyne species is an essential intermediate for aromatic and aliphatic amine.

**4.3. Role of Oxygen in the Formation of Imines and Dehydrogenation Pathway.** In contrast to the results of Trenary and co-workers,[11] no imine-like species were observed in temperature-dependent XP spectra. An important difference could lie in the different preparations of the single crystals. In this work, argon sputtering and annealing up to 900 K were sufficient to remove any carbon residues, as verified by XPS. Trenary and co-workers carried out argon sputtering and heating at 875 K in an oxygen atmosphere for 1 h to clean the surface, followed by an oxygen TPD up to 1000 K to rule out carbon residues remaining at the surface. However, this cleaning procedure may have resulted in the formation of a subsurface oxygen species. Subsurface oxygen can be formed at temperatures above 800 K and by long oxygen exposure but only desorbs at temperatures above 1250 K.[55,56] Besides this possibility, the formation of a chemisorbed oxygen species is also possible. The chemisorbed species would desorb at temperatures between 600 and 1100 K, but the occurrence of this species was also observed in the presence of a high concentration of subsurface oxygen.[56] Otherwise, the formation of chemisorbed oxygen is also possible through the adsorption of oxygen present in the residual gas of the chamber after the cleaning procedure. Note that oxygen is known to dissociate even below 100 K on Pt(111).[57]

Both subsurface and chemisorbed oxygen could influence the surface chemistry of NMA. To elucidate the influence, we dosed oxygen through the pinhole doser after adsorption of NMA, and the crystal was heated to 140 and 300 K (Figure 26). The N 1s detailed spectra show that two downshifted species develop at 140 K, which can be assigned to a deprotonated imine (398.1 eV)[58,59] and nitrile species (396.6 eV).[60,61] No significant changes occurred with further heating. The presence of N 1s species shifted to lower binding energies indicates a dehydrogenation of the amine moiety. Due to the low temperatures in the experiment, the role of subsurface oxygen can be ruled out. This experiment shows that coadsorbed oxygen plays a key role in the formation of imines, probably by acting as a hydrogen acceptor. Coadsorbed oxygen is known to have a promoting effect on the activation of C−H and N−H bonds of amines and alcohols adsorbed at gold surfaces since chemisorbed oxygen acts as Brönsted base leading to enhanced activation of C−H and N−H bonds.[62] An influence of coadsorbed oxygen can be ruled out in experiments shown here, as no oxygen was detected by XPS. Thus, the lack of formation of imines, such as observed in the experiments by Trenary and co-workers, can be attributed to missing oxygen species in the experiments presented here.

## 5. CONCLUSIONS

To summarize, the theoretical and experimental results presented above show that at the considered low coverage, the NMA molecule adsorbs with its phenyl ring parallel to the Pt(111) surface and the nitrogen atom at the top of a Pt surface atom, with the formation of a strong chemisorption bond. The most preferred adsorption site has the center of mass of the phenyl ring at the bridge site, with a bridge (as shown in Figure 2c) orientation of the phenyl ring. This relaxed geometry of the phenyl ring has two notable features: the elongation of the C−C bonds and the lifting up of the H atoms from the plane of the ring. These are due to the strong interaction of the phenyl ring of the molecule with the surface. We conclude that this bridge orientation gains extra stability because of the interaction between the Pt and C atoms, involving two C atoms and one Pt atom. At high coverages, NMA adsorbs with the nitrogen atom at the top site, and the phenyl ring inclined at an angle with respect to the surface. This phenyl ring tilting at the higher coverage is the result of a lateral repulsive interaction among the neighboring molecules.

The coverage dependence of the characteristics of adsorption of NMA on Pt(111) displays very interesting consequences experimentally in the XPS and FTIR data where it is observed that the molecule desorbs intact from the surface at a monolayer coverage and dissociates into fragments at sub-monolayer coverage. Bader charge analysis shows that the molecule shares charge through the phenyl ring and nitrogen atom at low coverage, whereas at the high coverage, charge sharing is through the nitrogen atom only. Furthermore, our calculations reveal that the formation of a phenylaminocarbyne species is a key intermediate step in the decomposition process, leading to the formation of hydrogen cyanide. There is a preference for C−H bond cleavage, and the cleavage of the C−N bond follows the activation of the C−H and N−H bonds. The results obtained here enhance our understanding of factors influencing bond activation pathways, which are crucial for designing new catalytic processes for amine reactions on metal surfaces.

## ■ ASSOCIATED CONTENT

### Ⓢ Supporting Information

The Supporting Information is available free of charge at https://pubs.acs.org/doi/10.1021/acs.jpcc.4c08116.

Theoretical data: Electronic band structure of Pt bulk (Figure S1); adsorption sites of the Pt(111) surface (Figure S2); PDOS of the Pt(111) surface and the molecule in the gas phase (Figures S3 and S4); minimum adsorption configuration of the molecule at the low coverage (Figure S5); coverage-dependent evolution of the geometrical structure of the adsorbed molecule on the Pt(111) surface (Figure S6); coverage-dependent PDOS of the adsorbed NMA molecule on the Pt(111) surface (Figures S7−11); charge density difference plots of the high, intermediate and low coverage systems (Figure S12) and Bader analysis (Tables S1−2); vibrational frequencies of the molecule in gas phase and adsorbed on the Pt(111) surface (Figures S13−20 and Table S3); coverage-dependent variation in the bond lengths of each atom in the molecule (Table S4); top view of 1/6 coverage with labeled atoms (Figure S21); top view of 1/36 coverage with labeled atoms (Figure S22); Experimental data: List of mass fragments for N-methylaniline (Table S5); temperature-programmed desorption spectra of an adsorbed multilayer N-methylaniline at Pt (111) (Figure S23); graphical analysis showing results of Redhead method (Figure S24); O 1s XP spectrum after dosing procedure (Figure S25); and XP spectra of an adsorbed multilayer of N-methylaniline coadsorbed with oxygen on Pt (111) (Figure S26) (PDF)

## ■ AUTHOR INFORMATION

### Corresponding Authors

Katharina Al-Shamery − *Institute of Chemistry, Carl von Ossietzky University of Oldenburg, 26129 Oldenburg, Germany;* orcid.org/0000-0002-4716-3240; Email: katharina.al.shamery@uni-oldenburg.de

Talat S. Rahman − *Department of Physics, University of Central Florida, Orlando, Florida 32816, United States;* Email: talat@ucf.edu

### Authors

Bushra Ashraf − *Department of Physics, University of Central Florida, Orlando, Florida 32816, United States*

Nils Brinkmann − *Institute of Chemistry, Carl von Ossietzky University of Oldenburg, 26129 Oldenburg, Germany;* orcid.org/0000-0002-2131-3773

Dave Austin − *Department of Physics, University of Central Florida, Orlando, Florida 32816, United States*

Duy Le − *Department of Physics, University of Central Florida, Orlando, Florida 32816, United States;* orcid.org/0000-0001-6391-8757

Complete contact information is available at:
https://pubs.acs.org/10.1021/acs.jpcc.4c08116

### Author Contributions

§B.A. and N.B. contributed equally to this work.

### Notes

The authors declare no competing financial interest.

## ■ ACKNOWLEDGMENTS

The computational work (B.A., D.A., D.L., T.S.R.) was supported by the US National Science Foundation grant CHE-1955343 and CHE-2400068. The DFG is thanked for their funding of the XPS system [INST 184/209-1 FUGG].

Nils Brinkmann was financially supported by the DFG (RTG 2226 "Chemical Bond Activation") which is greatly acknowledged. This work benefitted from the Helene Lange Award given to TSR by the University of Oldenburg.